\documentclass[twocolumn,nofootinbib,amsmath,amssymb,prl,aps,superscriptaddress,tightenlines,preprintnumbers]{revtex4}

\usepackage{hyperref}
\usepackage{graphicx}
\usepackage{graphics}
\usepackage{braket}
\usepackage[usenames]{color}
\newcommand{\be}{\begin{equation}}
\newcommand{\ee}{\end{equation}}
\newcommand{\bea}{\begin{eqnarray}}
\newcommand{\eea}{\end{eqnarray}}
\newcommand{\ba}{\begin{array}}
\newcommand{\ea}{\end{array}}
\newcommand{\Lagr}{\mathcal{L}}

\newcommand{\nn}{\nonumber}

\def\nn{\nonumber\\ }

\def\hyp{\mathsf{y}}

\begin{document}

\newcount\hour \newcount\minute
\hour=\time \divide \hour by 60
\minute=\time
\count99=\hour \multiply \count99 by -60 \advance \minute by \count99
\newcommand{\mydate}{\ \today \ - \number\hour :00}

\title{A methodology for theory uncertainties in the SMEFT}

\author{
Michael Trott${}^1$\\
${}^1$ Niels Bohr Institute, University of Copenhagen,
Blegdamsvej 17, DK-2100, Copenhagen, Denmark}

\begin{abstract}
A process specific methodology is defined to systematically assign theoretical uncertainties in the
Standard Model Effective Field Theory when performing leading order global fits. The method outlined
also minimizes the computational and theoretical burden to systematically advance such analyses to dimension eight.
 \end{abstract}

\maketitle
\newpage

\paragraph{\bf I. Introduction:}
The Standard Model (SM) is an incomplete description of observed phenomena in nature,
and usefully thought of as an Effective Field Theory (EFT)
for data analysis with characteristic energies around the Electroweak scale:
$\bar{v}_T \equiv \sqrt{2 \, \langle H^\dagger H \rangle}$.

The Standard Model Effective Field Theory (SMEFT) is based on the infared
assumptions that physics beyond the SM
is present at scales $\Lambda >\bar{v}_T$,
that there are no light hidden states in the spectrum with couplings
to the SM, and a $\rm SU(2)_L$ scalar doublet ($H$) with hypercharge
$\hyp_h = 1/2$ is present in the EFT.
A power counting expansion in
the ratio of scales $q^2/\Lambda^2 <1$, with $q^2$ a kinematic invariant associated
with experimental measurements in the domain of validity of the EFT, defines the SMEFT Lagrangian
\begin{align}
	\Lagr_{\textrm{SMEFT}} &= \Lagr_{\textrm{SM}} + \Lagr^{(5)}+\Lagr^{(6)} +
	\Lagr^{(7)} + \dots,  \\ \nonumber \Lagr^{(d)} &= \sum_i \frac{C_i^{(d)}}{\Lambda^{d-4}}\mathcal{Q}_i^{(d)}
	\quad \textrm{ for } d>4.
\end{align}
Higher dimensional operators ($\mathcal{Q}_i^{(d)}$) define SMEFT corrections to the SM predictions,
and carry a mass dimension $d$ superscript.The notation is such that
$\tilde{C}_{i}^{(6)} = \bar{v}_T^2 \, C_i^{(6)}/\Lambda^2$ and $\tilde{C}_{i}^{(8)} = \bar{v}_T^4 \, C^{(8)}_i/\Lambda^4$.
The operators multiply dimensionless Wilson coefficients $C_i^{(d)}$, which take on specific
values as a result of the Taylor expanded effects of physics beyond the SM. As the nature of physics beyond the SM is unknown,
we treat these Wilson coefficients as free parameters to constrain from the data.
The sum over $i$, after non-redundant operators are removed with field redefinitions
of the SM fields, runs over the operators in a particular operator basis.
We use the Warsaw basis \cite{Buchmuller:1985jz,Grzadkowski:2010es} for $\Lagr^{(6)}$ in this paper.
For higher order operators of mass dimension greater than six,
we use the conventions in Refs.~\cite{Helset:2018fgq,Corbett:2019cwl,Helset:2020yio,Hays:2020scx},
which defines the geoSMEFT formulation of this theory to all orders in $\sqrt{2 \langle H^\dagger H\rangle/\Lambda}$
for $n$-point functions, with $n\leq 3$.

Using the SMEFT, ATLAS and CMS have started to perform global analysis
of LHC data. Such global fits are performed in the context of theoretical predictions
of limited precision in the SMEFT. As approximations are used in the
theory predictions, it is required to define a theoretical error in SMEFT studies, to avoid miss-interpreting
experimental results in global SMEFT studies. In this paper we define such a methodology.

\paragraph{\bf II. Missing higher order terms:}
When a prediction is made in the SMEFT at leading order (LO), sub-leading terms are neglected.
The expansions present in the theory are
\begin{itemize}
\item{the {\textit{loop expansion}} in $g^2_{SM}/16 \pi^2$, where $g_{SM} \subset \left[g_1,g_2,g_3,\lambda, y_\psi \right]$
and $y_\psi$ is the Yukawa coupling for the fermion species $\psi$;}
\item{the {\textit{vev expansion}} in $\sqrt{2 \langle H^\dagger H \rangle}/\Lambda$, where $\langle H^\dagger H \rangle$
is the vacuum expectation of the Higgs field in the SMEFT; this expansion is relevant when SM kinematics
is present in a process due to a resonant SM state fixing $q^2 \simeq \bar{v}_T^2$, and}
\item{the {\textit{derivative expansion}} in $q^2/\Lambda^2$.}
\end{itemize}
Collectively we refer to the $\bar{v}_T/\Lambda$ and $q^2/\Lambda^2$ expansion as the SMEFT operator expansion.\footnote{
The expansion in $\bar{v}_T/\Lambda$ is relevant if some SM particles go on shell or are nearly on shell in an observable.
We refer to these observables as pole observables in this work. Conversely the expansion in  $q^2/\Lambda^2$
is relevant when considering non-resonant regions of phase space. We refer to such observables as tail observables in this work.}
There are also cross-terms in these expansions.  LO in the SMEFT means considering
$\mathcal{L}^{(6)}$ perturbations to the SM predictions, and roughly $\sim 30$ new parameters
impact Higgs, electroweak, and top-quark processes \cite{Brivio:2017btx}. Global SMEFT fits seek to constrain
these parameters. Uncertainties due to the truncation of the EFT expansion
due to missing higher-order terms should be assigned in this effort, if the historically standard methodology
of EFT studies of experimental data\footnote{See for example Refs.~\cite{nuc,Pich:1995bw,Bauer:2004ve,Abbate:2010xh,Passarino:2012cb,David:2013gaa,Dobaczewski:2014jga,Carlsson:2015vda,Hill:2010yb}.}
is to be followed in the case of SMEFT studies of LHC data.

At LO, a predicted (dimensionless SM) amplitude includes a perturbation due to $\Lagr^{(6)}$.
We illustrate the methodology with a pole observable where
\bea
\mathcal{A} = \mathcal{A}_{SM} + \tilde{C}^{(6)}_i \, a_i  + \cdots
\eea
here $a_i$ is a numerical coefficient that is process dependent. Sums over repeated indicies are implied.
The expression is not exact and is only defined to order $\mathcal{O}(1/\Lambda^2)$. It should be understood
to have arbitrary and unfixed
corrections of order $\mathcal{O}(1/\Lambda^4)$, until the SMEFT corrections for this process is defined
to order $\mathcal{O}(1/\Lambda^4)$. We return to this point in Section {\bf V}.
Quantum corrections
cannot be forbidden as the SMEFT is built of the SM fields.
As higher order terms in the loop and operator expansion are unknown, but necessarily present, it is important
to access the impact of neglecting these terms in LO SMEFT analyses of data.
The next order in the missing terms in both expansions are
\bea
\mathcal{A} &=& \mathcal{A}_{SM} + \tilde{C}^{(6)}_i \, a_i
+ \tilde{C}^{(6)}_j \, \tilde{C}^{(6)}_k \, b_{jk}
+ \tilde{C}^{(8)}_l \, c_{l}, \nn
&+& \frac{1}{16 \pi^2}
\left[d_{m} \, \tilde{C}^{(6)}_m + e_{n}\, \tilde{C}^{(6)}_n \, \log \left(\frac{\mu^2}{\Lambda^2}\right)\right]  +\cdots
\eea
The expression is also not exact and $a_i,b_{jk},c_l,d_m,e_n$ are numerical coefficients
that are process dependent. In each case, the indicies $i,j,k \cdots$ run over a subset of the full set of operators
in $\mathcal{L}^{(6)}$ and/or $\mathcal{L}^{(8)}$.

Squaring this expression, integrating over phase space, with relevant experimental cuts one finds
a cross section that is (schematically)
\bea\label{neglectedcorrections}
\sigma_{1/\Lambda^{4}} &=& \sigma_{SM} + \sum A_i \tilde{C}^{(6)}_i
 + \sum B_{jk} \tilde{C}^{(6)}_j \, \tilde{C}^{(6)}_k , \nn
&+& \sum D_l\tilde{C}^{(8)}_l+ \cdots
\eea
We assume that a LO simulation, including relevant experimental cuts and acceptance
corrections of the SMEFT are known for an observable, so the $A_i$ are fully known
for all $i$ in $\mathcal{L}^{(6)}$ in each process using a SMEFTsim \cite{Brivio:2017btx,Brivio:2020onw} based simulation.
Such results can also be produced directly for a class of terms in $B_{jk}$ with operators residing on different vertices,
for discussion see Ref.~\cite{Brivio:2020onw}. A SMEFTsim based simulation does not produce
the effects of canonically normalizing the SMEFT to $\mathcal{O}(1/\Lambda^4)$ and $\mathcal{L}^{(8)}$ operators.

To estimate the effect of the error due to neglecting higher order terms for $\sigma$,
knowing the actual prediction of all terms at next order in the loop expansion, and operator expansion,
and varying the unknown parameters in the higher order terms is a well defined and straightforward procedure.
As such, we focus on how to directly extract sub-leading terms from the results of a LO simulation.

\paragraph{\bf III. Missing perturbative corrections:}
Higher order terms in the SMEFT expansion that have common kinematic populations of phase
space as operators at $\mathcal{L}^{(6)}$ (already in a LO simulation) will receive common numerical corrections
due to Monte Carlo event generation, and phase space/acceptance cuts in an observable. This means that
for classes of $\mathcal{L}^{(8)}$, $(\mathcal{L}^{(6)})^2$ and $\mathcal{L}^{(6)}/16 \pi^2$
terms, these terms can be produced with appropriate rescaling from LO simulation results,
without the need of redundant and costly Monte-Carlo event generation.
The key to the methodology we lay out here is to define the appropriate approach to rescaling to leverage this
fact in practice.

Consider a partonic scattering process $X \rightarrow Y$.
This (implicitly) defines a tree-level SMEFT amplitude $\mathcal{A}_{X \rightarrow Y}(C_i/\Lambda^2)$ that
interferes with the SM amplitude $\mathcal{A}^{SM}_{X \rightarrow Y}$, the later of which can be of
any perturbative order. By definition
\begin{equation}
\sum A_i \tilde{C}^{(6)}_i \,  \propto \mathcal{A}^{SM}_{X \rightarrow Y} \times \mathcal{A}_{X \rightarrow Y}(\tilde{C}^{(6)}_i),
\end{equation}
up to suppressed phase space integrals with cuts.

We estimate perturbative uncertainties on the EFT parameters $C^{(6)}_i$ using the general form for one-loop
perturbative corrections to the SMEFT amplitude
\begin{equation}
 \left[a_i \tilde{C}^{(6)}_i + \frac{d_{m} \tilde{C}^{(6)}_m}{16 \pi^2}+ \frac{e_{in}\tilde{C}^{(6)}_n}{16 \pi^2} \log \frac{\mu^2}{\Lambda^2}\right].
\end{equation}
The $d_{m}$ terms can be determined with a dedicated one-loop calculation, and are in general unknown.
The coefficients $e_{in}$ of the log-enhanced terms, which generally give the largest contribution
to the perturbative correction, are known. The subscript has explicit dependence on $i$,
the coefficient appearing in the LO simulation, as these corrections have associated divergences.
Such divergences are canceled by the one loop renormalization of the process of interest, and must feed
in via a tree level operator dependence. The relevant calculations were determined in the one-loop renormalization group
evolution (RGE) of the SMEFT in Refs.~\cite{Jenkins:2013zja,Jenkins:2013wua,Alonso:2013hga}.
A bare SMEFT Lagrangian $\mathcal{L}^{(6),0}$ is coded into SMEFTsim and it is related
to the renormalized Lagrangian (denoted with an $(r)$ superscript)
where the SMEFT RGE counterterms $Z_{i,j}$ are introduced via
\begin{equation}
\mathcal{L}^{(6),r} = Z_{SM} \, Z_{i,n} \frac{C_i^{(6)}}{\Lambda^2} \mathcal{Q}_n^{(6),r}
\end{equation}
This rescaling by $Z_{i,n}$ is again \textit{transparent to the simulation chain}.
It is UV physics that scales with the leading order Wilson coefficient dependence, that is known,
and being used in the fit. The divergences exactly cancel after renormalization, however
the log terms that are associated with the divergences, defining $e_{in}$ do not cancel, but are predicted.

The log terms have the interpretation, when retained, of the Wilson coefficients being fit at the scale
$\Lambda$ vs being fit to at the the measurement scale $\mu$. The typical momentum transfer of a process fixes $\mu^2$, e.g.
for inclusive on-shell Higgs-boson decay
$\mu^2 = m_h^2$.
If renormalization group improved perturbation theory
is used all of the log terms can be summed running between $\Lambda$ and $\mu$ using standard EFT techniques.
When doing a LO SMEFT fit in fixed order perturbation theory, not preforming any resummation,
the log terms are the (generally largest) part of the theoretical error due to neglected terms in perturbation theory.
The log terms are not the full perturbative correction. $Z_{SM}$ refers to the SM
Wavefunction/$\bar{v}_T$ renormalization above. Importantly, this causes extra log terms
in addition to those inferred with this procedure \cite{Hartmann:2015oia}.
It is important to note that $Z_{SM}$ is also modified with dependence on $C_i^{(6)}$. This dependence
is an example of operators mixing down, and is always proportional to $\lambda$.
This dependence is fully given in Ref.~\cite{Jenkins:2013zja}.
Such corrections can be included this into the error estimate, the procedure is the same, one just
rescales the SM couplings with these modifications of the running of the SM parameters.
The top Yukawa ($y_t$) and gauge coupling ($g_3$) terms is expected
to generally introduce the dominant theory error.
The RGE is insufficient to characterize the full perturbative corrections and use for determining central values
of parameters \cite{Hartmann:2015oia}; it does not give a particularly good
approximation for the full perturbative correction at lower values of $\Lambda$.
However, practically, for such lower values of $\Lambda$ the $\mathcal{L}^{(8)}$ corrections that are also neglected are
expected to dominate the error estimate. Using the log terms as a reasonable proxy for unknown perturbative corrections,
for the purpose of a perturbative theory error when $\Lambda \gg \bar{v}_T$ is sufficient,
well defined, and known at this time.

This procedure can be applied to all processes in the fit. It is not limited to processes proceeding through
low n-point interactions. It is known that
to preserve the Ward Identites of the SMEFT \cite{Corbett:2019cwl} operator by
operator, it is necessary to expand out the propagator shifts in the SM masses
due to higher dimensional operators \cite{Corbett:2020ymv}.
These linear mass shifts (in particular to the $W$ mass in the case of an $\alpha$ input parameter scheme)
should be considered part of the theoretical prediction at leading order.
The perturbative error algorithm can be applied to such corrections. On the other hand the shifts
in the width are not required to preserve the Ward ID at one loop and the pert error algorthm should not be applied
to such terms. This is due to the one loop nature of the widths of the SM particles.
Similarly, if a process has already been improved to one loop in the SMEFT amplitude,
this procedure should not be used as the equivalent two loop RGE should be used. Partial results of this form
are available \cite{Bern:2020ikv}, but the full two loop-RGE remains unknown making this impractical to execute.

The log-enhanced one-loop correction to the SMEFT amplitude modifies each observable according to
\begin{equation}
\nabla \sigma_{1/16 \pi^2 \, \Lambda^2}  \approx
        \sum A_i \tilde{C}_j \frac{e_{ij}}{16 \pi^2} \log \frac{\mu^2}{\Lambda^2},
\end{equation}
and provides an estimate of the neglected perturbative corrections.
We reserve the notation $\nabla$ for errors in this work.

\subsection{Example: VBF Higgs production}
To demonstrate the determination of perturbative uncertainties we consider the inclusive result for the
``$\bar{q} q \rightarrow h \bar{q} q$ VBF-like'' 
process quoted in Table 9 of Ref.~\cite{Brivio:2020onw} (the ``direct'' contributions).
In this case,
\begin{equation}
\frac{A^{h \bar{q} q \, VBF-like}_i}{\sigma^{h \bar{q} q \, VBF-like}_{SM}} = \{-6,0.109,-5.345,-0.323,0.103,3\},
\end{equation}
for $\tilde{C}_i^{(6)} = \{\tilde{C}_{H \ell}^{(3)},\tilde{C}_{H q}^{(1)},\tilde{C}_{H q}^{(3)},\tilde{C}_{H u},\tilde{C}_{H d},\tilde{C}_{\ell \ell}'\}$.
For simplicity, we restrict our attention to perturbative corrections proportional to the top quark Yukawa in this example,
these results are present in Ref.\cite{Jenkins:2013wua}; specifically in Eqns.(A.33, A.27,A.28,A.30,A.29,A.35) of this work.
By inspection
\begin{equation}
\tilde{C}_j^{(6)} e_{ij} = - 2 \, N_c \, \tilde{C}^{(3)}_{\substack{\ell q \\ pp33}} y_t^2 + 2 \, N_c \, y_t^2 \tilde{C}_{\substack{H \ell \\pp}}^{(3)} + \cdots
\end{equation}
for  $\tilde{C}_i^{(6)} = \tilde{C}_{H \ell}^{(3)}$. The perturbative error that follows is
\bea
&\sim& A^I_i \tilde{C}_j  \frac{e_{ij}}{16 \pi^2} \log \frac{\mu^2}{\Lambda^2}, \\
&\sim& - 6 \,
\frac{\sigma^{h \bar{q} q}_{VBF-like}}{16 \pi^2} \left[- 2 \, N_c \, \tilde{C}^{(3)}_{\substack{\ell q \\ pp33}} y_t^2 + 2 \, N_c \, y_t^2 \tilde{C}_{\substack{H \ell \\pp}}^{(3)} \right]
\, \log \frac{\mu^2}{\Lambda^2}. \nonumber
\eea
These results from Ref.~\cite{Brivio:2020onw} have the cuts $m_{jj} > 350 \, {GeV}$, $p_T(h)< 200 \, {GeV}$.
A reasonable kinematic invariant to choose for the $\mu$ dependence in the logarithm
is $\mu \sim 200 \, {\textrm GeV}$.
$p$ is a flavour label. Consistent with the lepton flavour assumption it is summed over $p = \{1,2,3\}$.
This procedure is repeated for all of
$\tilde{C}^{(6)}_i = \{\tilde{C}_{H \ell}^{(3)},\tilde{C}_{H q}^{(1)},\tilde{C}_{H q}^{(3)},\tilde{C}_{H u},\tilde{C}_{H d},\tilde{C}_{\ell \ell}'\}$
for each SM coupling dependence that one wishes to retain. For practical purposes retaining
the dependence on $y_t,y_b, g_{1,2,3}$ is sufficient. What results is an error estimate
that is a linear sum of unknown (nuisance) parameters, with calculated coefficients.
It is dominated in its SM coupling dependence by $y_t,g_3$ corrections.
A distribution of the unknown Wilson coefficients needs to be chosen to produce a number to add in quadrature with other errors.
A very weak dependence is present on the distribution chosen (other than the overall scale $\Lambda$ choice), consistent
with the \href{http://www.medicine.mcgill.ca/epidemiology/hanley/bios601/GaussianModel/HistoryCentralLimitTheorem.pdf}{central limit theorem}.
This statistical behavior is present in the results shown in Refs.~\cite{Hays:2020scx,Corbett:2021eux}.

\paragraph{\bf III. Missing $\mathcal{O}(1/\Lambda^4)$ corrections:}
Many of the $\mathcal{O}(1/\Lambda^4)$ corrections can also be determined
from LO SMEFT results with re-scalings. It is appropriate to organize the theory in terms of specific composite operator
kinematics, with scalar dressings that do not introduce new kinematics to identify these rescalings.
This is exactly the geoSMEFT approach developed
in Refs.~\cite{Helset:2018fgq,Corbett:2019cwl,Helset:2020yio,Hays:2020scx,Corbett:2020bqv}.
In this case, field-space connections $G_i$
multiplying composite operator forms $f_i$ as
\bea\label{basicdecomposition}
\Lagr_{\textrm{SMEFT}} = \sum_i G_i(I,A,\phi \dots) \, f_i ,
\eea
where $G_i$ depend on the group indices $I,A$ of the (non-spacetime) symmetry groups,
and the scalar field coordinates of the composite operators. Powers of $D^\mu H$ are included in $f_i$.
The kinematic dependence is thus factorized into the $f_i$ and the rescalings by $G_i$
are exactly the re-scalings needed to produce $\mathcal{O}(1/\Lambda^4)$ corrections from LO simulation results.

The geoSMEFT has been defined for all interactions up to four point interactions at this time. The rescaling procedure is best illustrated
with a specific example. The three-point function $h-\gamma-\gamma$ in the SMEFT to all orders is given in Ref.~\cite{Helset:2020yio}.
$\langle h \gamma^{\mu\nu} \gamma_{\mu\nu}\rangle$ is a common kinematic factor
for the $\mathcal{L}^{(6,8)}$ contributions. As such, simple replacements can be made on the $\mathcal{L}^{(6)}$
dependence determined in SMEFTsim to directly generate $\mathcal{O}(1/\Lambda^4)$ terms from LO results.
Using the all orders definition of the decay width Ref.~\cite{Helset:2020yio,Hays:2020scx}
at LO one finds
\begin{align}
\frac{\Gamma^{\hat{m}_{W}}(h \rightarrow \gamma \gamma)}{\Gamma^{\hat{m}_{W}}_{\rm SM}(h \rightarrow \gamma \gamma)}
=1 -  788 f^{\hat{m}_W}_1,
\end{align}
Here $f^{\hat{m}_W}_i \simeq f^{\hat{\alpha}_{ew}}_i$ for $i=1,2,3$ and
\begin{align}
 \delta G_F^{(6)} &= \frac{1}{\sqrt2} \left(\tilde C^{(3)}_{\substack{Hl \\ee}}+\tilde C^{(3)}_{\substack{Hl \\ \mu \mu}} - \frac{1}{2}(\tilde C'_{\substack{ll \\ \mu ee \mu}}+\tilde C'_{\substack{ll \\ e \mu \mu e}})\right),\\
f^{\hat{m}_W}_1 &=  \left[\tilde{C}_{HB}^{(6)} +0.29 \, \, \tilde{C}_{HW}^{(6)} -0.54  \, \tilde{C}_{HWB}^{(6)}\right],\\
f^{\hat{m}_W}_2 &=   \left[\tilde{C}_{HB}^{(8)} +0.29 \, \, (\tilde{C}_{HW}^{(8)}+ \tilde{C}_{HW,2}^{(8)}) -0.54  \, \tilde{C}_{HWB}^{(8)}\right],\\
f^{\hat{m}_W}_3 &= \left[\tilde{C}_{HW}^{(6)} - \tilde{C}_{HB}^{(6)} -0.66  \, \tilde{C}_{HWB}^{(6)}\right],
\end{align}
and these expressions (approximately) hold in both input-parameter schemes.
The full result to $\mathcal{O}(1/\Lambda^4)$, in the $m_W$ scheme, is \cite{Hays:2020scx}
\begin{widetext}
\begin{align}
\frac{\Gamma^{\hat{m}_{W}}(h \rightarrow \gamma \gamma)}{\Gamma^{\hat{m}_{W}}_{\rm SM}(h \rightarrow \gamma \gamma)}
=1 &-  788 f^{\hat{m}_W}_1 + 394^2 \, (f^{\hat{m}_W}_1)^2
- 351 \, (\tilde{C}_{HW}^{(6)} - \tilde{C}_{HB}^{(6)})\, f^{\hat{m}_W}_3 + 2228 \, \delta G_F^{(6)} \, f^{\hat{m}_W}_1, \nonumber \\
&+  979 \, \tilde{C}_{HD}^{(6)}(\tilde{C}_{HB}^{(6)} +0.80\, \, \tilde{C}_{HW}^{(6)} -1.02  \, \tilde{C}_{HWB}^{(6)})
-788 \left[ \left(\tilde C_{H\Box}^{(6)} - \frac{\tilde C_{HD}^{(6)}}{4}\right)  \, f^{\hat{m}_W}_1+ f^{\hat{m}_W}_2\right], \nonumber \\
&+2283 \, \tilde{C}_{HWB}^{(6)}(\tilde{C}_{HB}^{(6)} +0.66 \, \, \tilde{C}_{HW}^{(6)} -0.88  \, \tilde{C}_{HWB}^{(6)})
- 1224 \, (f^{\hat{m}_W}_1)^2,
\end{align}
\end{widetext}
A leading order simulation result using SMEFTsim that gives dependence on the Wilson coefficients
in $f^{\hat{m}_W}_1$ or $f^{\hat{\alpha}_{ew}}_1$ (i.e. $\tilde{C}_{HB}^{(6)},\tilde{C}_{HW}^{(6)}, \tilde{C}_{HWB}^{(6)}$),
with numerical dependence multiplying this function in an observable $N_i$
can then be used to generate the remaining $\mathcal{O}(1/\Lambda^4)$ terms with the replacement
\begin{widetext}
\begin{align}
N_i \, f^{\hat{m}_W}_1 & \rightarrow \frac{-N_i}{788} \left[ 394^2 \, (f^{\hat{m}_W}_1)^2
- 351 \, (\tilde{C}_{HW}^{(6)} - \tilde{C}_{HB}^{(6)})\, f^{\hat{m}_W}_3 + 2228 \, \delta G_F^{(6)} \, f^{\hat{m}_W}_1, \right. \nonumber \\
&+  979 \, \tilde{C}_{HD}^{(6)}(\tilde{C}_{HB}^{(6)} +0.80\, \, \tilde{C}_{HW}^{(6)} -1.02  \, \tilde{C}_{HWB}^{(6)})
-788 \left[ \left(\tilde C_{H\Box}^{(6)} - \frac{\tilde C_{HD}^{(6)}}{4}\right)  \, f^{\hat{m}_W}_1+ f^{\hat{m}_W}_2\right], \nonumber \\
& \left.+2283 \, \tilde{C}_{HWB}^{(6)}(\tilde{C}_{HB}^{(6)} +0.66 \, \, \tilde{C}_{HW}^{(6)} -0.88  \, \tilde{C}_{HWB}^{(6)})
- 1224 \, (f^{\hat{m}_W}_1)^2
\right].
\end{align}
\end{widetext}
The use of this replacement in the one loop result for the $f^{\hat{m}_W}_1$ dependence in $\Gamma(h \rightarrow \gamma \gamma)$
introduces a relative uncertainty of $(\bar{v}_T^4/\Lambda^4)(1/16 \pi^2)$. The replacement generates not only the
quadratic terms, but also the full set of $\bar{v}_T^4/\Lambda^4$ correction contributing to
$\nabla \Gamma(h \rightarrow \gamma \gamma)$.
An error can then be assigned by choosing
a set of distributions for the $C_i^{(6)},C_i^{(8)}$ and a value for $\Lambda$ when neglecting this class of terms.
The choice of $\Lambda$ dictates the size of the error induced, and it is appropriate to choose multiple values
of $\Lambda$ when determining errors. A straightforward choice is to choose $\Lambda \sim 1 \, {\rm TeV}$ and
$\Lambda \sim 3 \, {\rm TeV}$. $\mathcal{L}^{(8)}$ induced errors dominate for the former choice, while errors due to neglected
perturbative corrections dominate for the later choice.

The case of $h-\gamma-\gamma$ is particularly simple, due to a narrow width approximation
for Higgs factorising production and decay, and the presence of only one vertex with a common kinematic structure.
Extending this procedure to processes where multiple Feynman diagrams contribute, where each vertex building up the
individual Feynman diagrams is generalised into the case of the geoSMEFT, requires the individual dependence on at least
one Wilson coefficient present at $\mathcal{L}^{(6)}$ in each type of vertex be identified and isolated, so that a rescaling procedure using
the geoSMEFT generalisation of that vertex can be performed. As the same $\mathcal{L}^{(6)}$ correction can appear
in multiple vertices, this can require choosing combinations of Wilson coefficients
with fixed linear algebra relations to project out the dependence of a Wilson coefficient at a particular vertex.
For example, consider a process where the same $\mathcal{L}^{(6)}$  Wilson coefficient $C_1^{(6)}$ appears in two vertices
in a Feynman diagram, or sum of diagrams with dependences
\bea
\delta V_1 &\propto& a_1 \, \tilde{C}_1^{(6)}+ a_2 \, \tilde{C}_2^{(6)} + a_3 \, \tilde{C}_3^{(6)}, \nn
\delta V_2 &\propto& b_1 \, \tilde{C}_1^{(6)}+ b_2 \, \tilde{C}_2^{(6)} + b_3 \, \tilde{C}_3^{(6)}.
\eea
The kinematics associated with the vertices $\delta V_{1,2}$ can differ in what follows.\footnote{Here $a_{1,2,3}$ and $b_{1,2,3}$ refer
to explicit analytic (or numerically evaluated) dependence on a Wilson coefficient in a vertex. For example,
when $\delta V_1 = f_1^{\hat{m}_W}$ then $a_{HB} = 1, a_{HW} = 0.29, a_{HWB} = -0.54$.}
Each of the $V_1$ and $V_2$ have a rescaling in the geoSMEFT, but the appearance in the overall result
is a convolution of dependence on $\tilde{C}_{1,2,3}^{(6)}$ from both vertices. Isolating the dependence on $C_1^{(6)}$ in
$V_1$ to perform the required rescaling to generate the $\mathcal{O}(1/\Lambda^4)$ result, one can choose to fix $\delta V_2 = 0$
in the known SMEFT result by choosing $\tilde{C}_2^{(6)} = (- b_1 \,  \tilde{C}_1^{(6)} - b_3 \, \tilde{C}_3^{(6)})/b_2$ in the LO result.
The resulting shift due to $\delta V_1$ is then modified to
\bea\label{shiftedresult}
\delta V'_1 &\propto& \frac{a_1 - a_2 \,b_1}{b_2} \, \tilde{C}_1^{(6)} + \frac{a_3 -a_2 \, b_3}{b_2} \, \tilde{C}_3^{(6)}.
\eea
The geoSMEFT based rescaling uses the known dependence of $ \delta V_1 = a_1 \, \tilde{C}_1^{(6)}$
and a descendent $\mathcal{L}^{(8)}$ result $ \delta V_1 = b_{jk} \, \tilde{C}_j^{(6)} \, \tilde{C}_k^{(6)} + c_l \, \tilde{C}_l^{(8)}$.
The net dependence on $\tilde{C}_1^{(6)}$ in Eqn.~(\ref{shiftedresult}) can be rescaled back to a net $a_1$ dependence
using the known dependence on all vertices in the contribution to the observable; i.e. $a_1,a_2,b_1,b_2,b_3$ are all
known analytically in the LO SMEFT results encoded in SMEFTsim, that are consistent with the geoSMEFT generalisation to
higher orders in $1/\Lambda$. This procedure can be iterated for Wilson coefficient dependence in more than two vertices.
Performing the full set of rescaling replacements for all vertices that build up the Feynman diagrams contributing
to an observable then generalises the LO SMEFTsim result with a well defined class of terms at $\mathcal{O}(1/\Lambda^4)$.
A chosen set of distributions of $C_i^{(6)}$ and $C_i^{(8)}$, and far more importantly, a chosen $\Lambda$ scale,
then defines a numerical error for this set of calculated terms.

\paragraph{\bf IV. New kinematics beyond $\mathcal{L}^{(6)}$:}
The procedure outlined above, relies on the LO simulation for global SMEFT studies having
a complete set of kinematic functions to rescale. A relevant question is when new kinematic forms
appear first at $\mathcal{L}^{(8)}$ in SMEFT global fits of Higgs and EW data. Such anomalous kinematics
is remarkably limited when only considering $n\leq 3$ point interactions building up pole observables.
The anomalous kinematics are limited to the field space connections of the geoSMEFT of the form
\begin{equation}
k_{IJA}(\phi)(D^\mu \phi)^I(D^\nu\phi)^J \mathcal{W}_{\mu\nu}^A.
\end{equation}
The operator contributions to this field space connection
are equation of motion (EOM) reducible at $\mathcal{L}^{(6)}$, and hence are not present in the Warsaw basis.
At $\mathcal{L}^{(8)}$ such terms are no longer EOM reducible.
Such contributions modify $VV \rightarrow h$ production,
and $h \rightarrow 4 \ell$ through the modification of the $hVV$ vertex (here $V$ is a general vector).
These corrections must be added in a dedicated extension of SMEFTsim \cite{modSMEFTsim}.\footnote{This modification is available from the author and T. Corbett.} Similarly, modified kinematics is present in $VH$ production
which requires a dedicated extension of simulation tools, see Ref.~\cite{Hays:2018zze}. These contributions can be directly
targeted for code extensions and direct simulation to complete out the calculation of relevant observables to $\mathcal{O}(1/\Lambda^4)$
using the $\mathcal{L}^{(8}$ operator basis in Refs.~\cite{Murphy:2020rsh,Li:2020gnx}.
Such dedicated extensions to SMEFTsim by subsets of $\mathcal{L}^{(8)}$ operators, combined with the algorithm above results in a fully
well defined theoretical result to $\mathcal{O}(1/\Lambda^4)$ process by process, and such results can then be used
in order to generate a theory error by directly varying the sub-leading terms in a chosen manner.
The majority of the $\mathcal{O}(1/\Lambda^4)$ results, when considering pole observables,
can be generated for this purpose with geoSMEFT based rescaling,
avoiding the need for simulation or code modification. On the other hand, when considering tail observables,
four point interactions are generally important and unsuppressed compared to other SMEFT corrections.\footnote{See for example Ref.~\cite{Boughezal:2021tih}.}
It is then required to add more operators to
simulation tools to characterise observables to $\mathcal{O}(1/\Lambda^4)$. When considering four point
interactions (with no scalar field) a geoSMEFT based rescaling of lower order results
is not possible at this time.

\paragraph{\bf V. Quadratic terms:}\label{quadratics}

There is some confusion in the literature on the nature of quadratic terms.
We address this issue to assess the use of quadratic terms for a theory error
estimate, and if the use of quadratic terms for SMEFT fits to define central values
in the fit is well defined.

Here quadratic terms, means the result of squaring
Eqn.~(2); the resulting $\mathcal{O}(1/\Lambda^4)$ term is the quadratic term. On general grounds retaining only
a subset of terms in the power counting of an EFT is an ill-defined procedure, which is not invariant
under the field redefinitions that define the theory.\footnote{One can directly confirm
that inconsistent expansions in the power counting expansion among vertex fucntions
violate SMEFT Ward identities \cite{Corbett:2019cwl}.} In the case of the SMEFT retaining quadratic terms is subject to the following field redefinition
based ambiguities. Eqn.~(2) should be understood to have unspecified but existent terms of the form
\bea
\mathcal{A} = \mathcal{A}_{SM}\left(1 + n_i \frac{\bar{v}_T^4}{\Lambda^4} \right) + \tilde{C}^{(6)}_k \, a_k \, \left(1 + o_j \frac{\bar{v}_T^2}{\Lambda^2} \right)+ \cdots
\eea
There are also corrections of $\mathcal{O}(1/\Lambda^4)$ with dynamical fields of dimension four and two in each case.
The freedom to perform field definitions of $\mathcal{O}(1/\Lambda^4)$ is fundamental to defining
the SMEFT, and such an ambiguity is not fixed defining the theory to $\mathcal{O}(1/\Lambda^2)$.
Predictions proportional to $n_i$ and $o_j$ are ambiguous until the full set of corrections are defined at
$\mathcal{O}(1/\Lambda^4)$ when defining an operator basis for $\mathcal{L}^{(8)}$.
Squaring this result gives terms of $\mathcal{O}(1/\Lambda^4)$
\bea\label{ambiguity}
\left(a_k \tilde{C}_k^{(6)} a_l \tilde{C}_l^{(6)} + 2  n_i \frac{\bar{v}_T^4}{\Lambda^4} |\mathcal{A}_{SM}|^2 + 2  a_k \mathcal{A}_{SM}  \tilde{C}^{(6)}_k  o_j \frac{\bar{v}_T^2}{\Lambda^2} \right)
\eea
All terms are of order $\mathcal{O}(1/\Lambda^4)$ and the $n_i$ and $o_j$ are arbitrary as it stands.
These terms can be chosen to have $\tilde{C}_i^{(6)}$ dependence in particular, modifying the dependence on
$\tilde{C}_k^{(6)} \tilde{C}_l^{(6)}$.

This arbitrariness at $\mathcal{O}(1/\Lambda^4)$  is represented by results in the published literature.
Ref.~\cite{Hays:2020scx} demonstrated that the leading quadratic term did not correctly
predict the dependence on $(f^{\hat{m}_W}_1)^2$.
The reason that the quadratic term does not correctly predict
the dependence on $(f^{\hat{m}_W}_1)^2$ is due to the redefinition of the electric coupling
in the geoSMEFT. This redefinition is a specific example of field redefinition based corrections of the form
shown in Eqn.~\eqref{ambiguity}.
In addition, the  arbitrariness represented by $n_i,o_j$ untill $\mathcal{L}^{(8)}$ is defined is required
to ensure that SMEFT predictions are well defined at $\mathcal{O}(1/\Lambda^4)$ and not intrinsically
dependent on the basis chosen for $\mathcal{L}^{(8)}$. When defining $\mathcal{L}^{(8)}$
the use of the Higgs EOM
\begin{align}
D^2 H_k &-\lambda \bar{v}_T^2 H_k +2 \lambda (H^\dagger H) H_k + \overline q^j\, Y_u^\dagger\, u \epsilon_{jk}, \nonumber \\
&+ \overline d\, Y_d\, q_k +\overline e\, Y_e\,  l_k =0.
\label{eomH}
\end{align}
leads to modifications of $\mathcal{L}^{(6)}$ terms that are $\propto \bar{v}_T^2$.
These terms are basis dependent artifacts that cancel in a full matching, as demonstrated in
Ref.~\cite{Hays:2020scx,Corbett:2021eux}. This correlated matching
of $\mathcal{L}^{(8)}$, and $ \bar{v}_T^2$ corrections to $\mathcal{L}^{(6)}$ is an example of matching
effects descending down in operator mass dimension.
This physics in matching and defining the operator bases, is similar to the
mixing of operators of different mass dimension in the SMEFT RGE; both effects come about due to
the presence of the classical scale $ \bar{v}_T^2$.
$\tilde{C}^{(6)}_k \, a_k$ is not well defined
in its predictions to $\mathcal{O}(1/\Lambda^4)$ when neglecting such effects,
as the Higgs EOM is not consistently applied at this order neglecting such terms.
The quadratic terms are in general not well defined contributions at order $\mathcal{O}(1/\Lambda^4)$ for these reasons,
they should not be used to fix central values in the global SMEFT fit as a default prediction.
Such corrections also cannot be translated unambiguously between operator bases
until the theory is fixed to $\mathcal{O}(1/\Lambda^4)$. The reason is that Eqn.~\eqref{eomH}, and other EOM relations between
$\mathcal{L}^{(6)}$ operators
have implicit (generally unspecified) corrections that are further suppressed by
$\mathcal{O}(1/\Lambda^2)$.\footnote{Such corrections to the SMEFT EOM are reported in Ref.~\cite{Barzinji:2018xvu}.}

Nevertheless, the use of quadratic terms to define an error estimate
is a reasonable procedure \cite{Berthier:2015gja,Alte:2018xgc,Keilmann:2019cbp,Alte:2017pme}, in the absence of
complete results developed using the method outlined here. In particular,
as the methodology outlined here is focused on improving SMEFT results to $\mathcal{O}(1/\Lambda^4)$
for pole observables, the use of quadratic terms to estimate an error for tail
observables as advocated in Ref.~\cite{Alte:2018xgc,Keilmann:2019cbp,Alte:2017pme} can be appropriate.

\paragraph{\bf VI. Conclusions:}
We have defined a methodology to improve predictions for pole observables in the SMEFT
systematically to $\mathcal{O}(1/\Lambda^4)$ using geoSMEFT results.
When new kinematics is first present at $\mathcal{L}^{(8)}$,
modifications to code tools, and new simulation and event generation is required to complete
results to $\mathcal{O}(1/\Lambda^4)$. However such corrections are a small subset of
the full set of corrections extending predictions to $\mathcal{O}(1/\Lambda^4)$. This approach to improving LO results to sub-leading order in the operator
expansion relies on simple linear algebra, Taylor expansions of known closed form all orders
expressions in the geoSMEFT, and the known dependence on $\mathcal{L}^{(6)}$ encoded in SMEFTsim.
The approach outlined here can be combined with the approach of
Refs.~\cite{Berthier:2015gja,Alte:2018xgc,Keilmann:2019cbp,Alte:2017pme} for non pole observables.

Truncation errors result from taking the resulting exact expressions to $\mathcal{O}(1/\Lambda^4)$ and
varying the unknown higher order terms in a range of values. Then
$\sqrt{(\nabla \sigma_{1/\Lambda^4})^2+ (\nabla \sigma_{1/16 \pi^2 \Lambda^2})^2}$
defines a theory error estimate. As the operator expansion in the SMEFT
involves many terms at $\mathcal{O}(1/\Lambda^4)$, that are randomly chosen via distributions in linear sums,
the central limit theorem applies.
In global combinations, common values of $\Lambda$ should be chosen to define errors for observables.
The resulting SMEFT theory error for a LO fit is a Gaussian distributed
numerical value for each observable, with magnitude determined by the chosen $\Lambda$.
\\
\\
{\bf Acknowledgements}
M.T. thanks members of the ATLAS collaboration for useful scientific discussions and encouragement.
M.T. acknowledges support from the Villum Fund, project number 00010102.
We thank A. Biekoetter, T. Corbett and A. Martin for comments on the draft.
\\
\bibliography{bibliography}
\end{document}